# Controlling Fano resonance using the Geometrical Phase of light in spatially tailored waveguided plasmonic crystals

Subir K. Ray[†], Ankit K. Singh[†]*, Ajmal, Shubham Chandel, Partha Mitra, Nirmalya Ghosh

*Department of Physical Sciences,*

*Indian Institute of Science Education and Research (IISER) Kolkata.*

*Mohanpur 741246, India*

[†] The authors contributed equally

*Corresponding authors: aks13ip027@iiserkol.ac.in*

**Abstract**

Fano resonance exhibiting an asymmetric spectral line shape is a universal phenomenon observed in diverse physical systems. Here we experimentally establish a direct link between the spectral asymmetry parameter "$q$" and a physically realizable phase factor of interference between a continuum and a discrete mode that leads to Fano resonance. Using a specially designed metamaterial, namely waveguided plasmonic crystal with a spatially varying orientation axis of plasmonic grating, we demonstrate control on the spectral asymmetry of the Fano resonance through changes in the geometric phase of polarized light. In this scenario, the changes in the geometric phase for input left, and right circular polarized light arises due to varying orientation angle of the grating axis. The systematic changes in the geometric phase and the resulting $q$-parameter of Fano resonance is interpreted by an appropriate theoretical model connecting the two physical entities. The demonstrated control over the spectral line shape of Fano resonance achieved by tailoring geometric phase may open up novel routes for polarization - based photonic applications.

Fano resonance is a ubiquitous interference phenomenon observed in a wide range of physical system including atomic, molecular, optical, nuclear, solid-state as well as in classical optical systems [1-9]. In contrast to the conventional symmetric Lorentzian resonance profile, the Fano resonance is uniquely identified in the spectral line shape as an asymmetric spectral profile where an intensity maximum, is immediately followed by an intensity minimum due to destructive interference. The spectral asymmetricity emerges due to interference between a continuum mode and a discrete mode, which can be characterized by the asymmetry parameter "$q$" [1]. In the optical domain, the asymmetric Fano resonance has been intensively studied not only due to the fundamental interests but also because of its numerous potential applications in sensing, switching, lasing, filters, robust color display, nonlinear and slow-light devices and so forth [10-20]. In most of the mentioned applications, the ability to control and tune the interference effect with some external parameters is crucial. In quest of making a suitable structure for a particular application, various type of metal-dielectric structures like plasmonic oligomers, dolmen, split rings, waveguided plasmonic crystal, etc. have been proposed to observe the optical Fano resonances in the scattering or absorption profile [6, 10-20].

Despite the availability of voluminous literature on the application aspects of Fano resonance in optical systems, there still remain considerable interests in understanding the fundamental nature of this intriguing wave interference phenomenon through some physically meaningful and experimentally accessible parameters [7, 21]. On conceptual and practical grounds such studies are extremely valuable since these not only provide new insights and interpretation of the Fano resonance phenomenon, but it may also open up novel routes to control and modulate the interference effect and the resulting Fano resonance [7, 21]. In this regard, we have recently proposed a simple phenomenological model for the electric field scattered from a Fano resonant system and related the spectral asymmetry of the resonance profile (or the $q$-parameter) with a physical parameter namely, Fano phase shift between the continuum and the discrete mode [21]. Here, we show that the Fano phase shift between the continuum and the discrete mode is indeed a physically meaningful phase factor of light and that this phase factor can be controlled to modulate the Fano $q$-parameter by controlling some degree of freedom of light. We demonstrate this concept in a spatially tailored waveguided plasmonic crystal sample that exhibits Fano resonance due to the spectral interference of a spectrally broad plasmon mode and a narrow waveguide mode. In this particular scenario, the Fano phase factor and the resulting spectral asymmetry parameter is directly related to the Pancharatnam Berry geometric phase of light [22-30], which is systematically changed by varying the orientation axis of the gold grating in the waveguided plasmonic crystal sample. The polarization degree of freedom of light, specifically opposite circular polarization states of light is subsequently used to probe systematic changes in the Fano spectral line shape of the scattered light as the signature of varying Fano phase shift. The corresponding experimental results are interpreted using a suitable theoretical model, which establishes the link between the Fano phase shift (which is related to geometric phase here) and the asymmetry parameter ($q$) of Fano resonance. These results uncover an intriguing link between the Fano $q$-parameter and a physically accessible phase of interference and also open up an interesting avenue for control and manipulation of optical Fano resonance through a geometrical phase and polarization state of light.

**Theory**

We first develop a theoretical model that connects the Fano asymmetry parameter $q$ to an additional polarization-dependent geometric phase factor that may arise in inhomogeneous anisotropic Fano resonant systems [24-26]. In the general case, the scattered electric field from a

Fano resonant system can be written as the interference of a discrete complex Lorentzian with a continuum of amplitude "$B(\omega)$" [21]

$$\mathbf{E}_s \approx \frac{q-i}{\epsilon+i} + B(\omega) = \frac{\sqrt{q^2+1}e^{i\psi_f}}{\epsilon+i} + B(\omega) \tag{1}$$

Here, $\varepsilon = \varepsilon(\omega) = \frac{\omega-\omega_0}{(\gamma/2)}$, $\omega_0$ and $\gamma$ are the central frequency and the width of the discrete mode respectively, $\psi_f$ is called Fano phase factor. Assuming frequency independent continuum mode, the scattered intensity corresponding to Eq. 1 can be shown to have a spectral intensity component that exhibits characteristic asymmetric spectral line shape along with a Lorentzian background as [21]

$$I_s = |\mathbf{E}_s|^2 = \frac{(q+B\epsilon)^2}{\epsilon^2+1} + \frac{(1-B)^2}{\epsilon^2+1} \tag{2}$$

If one of the contributing mode (either discrete or the continuum) has a phase contribution of geometrical origin, then the resulting interference effect will get modulated by the additional geometric phases acquired by the particular mode. In such case one of the contributing modes acquire a geometrical phase ($\sigma\phi_g = \pm\phi_g$) for incident LCP/RCP light [23-26], depending on the helicity ($\sigma$, "+/-" is for left/right circular polarization) of the light incident on the Fano resonant scattering system. The origin of such geometrical phase is due to the geometric orientation of the scattering system with respect to incident polarized light. The scattered electric field from such a system can be written as:

$$\mathbf{E}_s \approx \frac{\sqrt{q^2+1}e^{i\psi_f}}{\epsilon+i} + Be^{\pm i\,\phi_g} \tag{3}$$

The corresponding spectral variation of the intensity for incident LCP/RCP light can be written as

$$I_s^{LCP/RCP}(\omega,\phi) = \frac{\left(B\epsilon+(q\cos\phi_g \mp \sin\phi_g)\right)^2}{\epsilon^2+1} + \frac{\left((\cos\phi_g \pm q\sin\phi_g)-B\right)^2}{\epsilon^2+1} \tag{4}$$

As evident from eq. 3 and 4, the geometric phase of light can modulate the Fano phase shift $\psi_f$ or the resulting asymmetry parameter $q$. The difference in the geometrical phase between left and right circular polarization thus leads to a contrast in the scattering spectral line shape for the two polarizations.

**Results and Experiment**

In order to probe the circular polarization dependence of the Fano line shape we took a one dimensional waveguided plasmonic crystal, a well-studied optical system for the Fano resonance in the extinction spectra [6, 9]. It consists of 2-fold symmetric one-dimensional gold grating deposited on a thin Indium Tin oxide (ITO) coated quartz substrate as shown in **Figure 1**. The Fano resonance in such system is observed as a result of interference between a broad dipolar plasmonic resonance of the gold grating and the waveguide modes excited in the ITO layer [6, 9]. It is to be noted that the plasmonic grating also acts as a diattenuating linear retarder for the field incident on the system showing a simultaneous differential/anisotropic amplitude as well as the phase response for orthogonal Eigen polarization states respectively [23], as shown later. The direction of the grating lines decides the orientation axis of the decomposed linear diattenuator and linear retarder element present in the system.

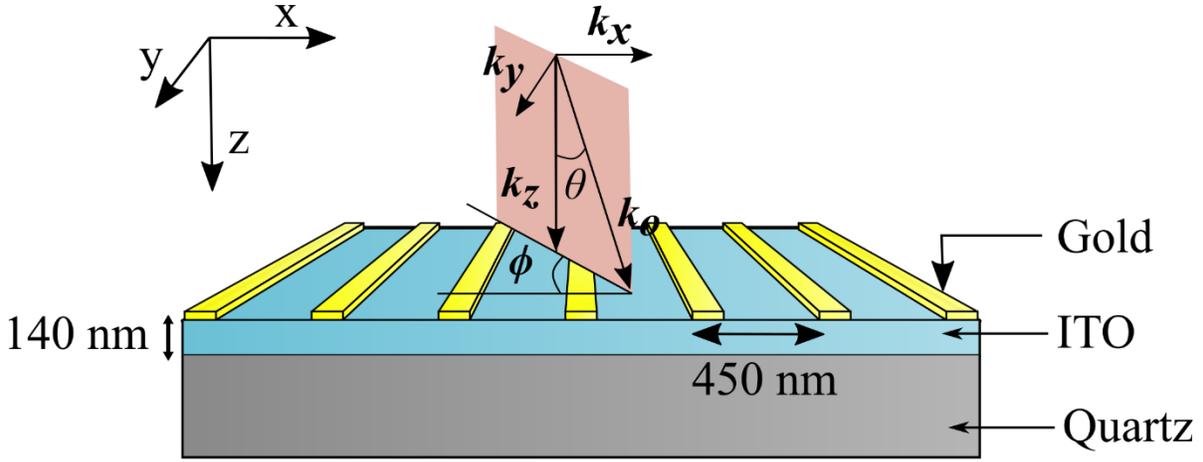

**Figure 1**: **A schematic of waveguided plasmonic grating crystal**. A gold grating of $100\ nm$ width and $20\ nm$ height on top of a $140\ nm$ thick Indium Tin Oxide (ITO) coated quartz substrate was taken for the finite element method based COMSOL Multiphysics. The extinction spectra were studied for light incident at an inclined angle of incidence ($\theta = 16°$) with varying azimuthal angle ($\phi$) of incidence from $0°$ to $90°$ (in steps of $15°$).

The extinction spectra (1-transmission) of a grating (height 20 nm width 100 nm and periodicity 450 nm) was calculated at an inclined angle ($\theta = 16^o$) with varying the azimuthal angle ($\phi$) of incidence from $0°$ to $90°$ in step of $15^o$, using finite element method (FEM) based COMSOL Multiphysics software. The extinction spectra were observed for both the circular polarized input states (i.e., LCP and RCP) with scattered field projected on the input as well as the orthogonal state of polarization as shown in **Figure 2**. It can be noted from the **Figure 2a** that for an inclined angle of incidence, a differential response of the system for incident LCP/RCP polarizations projected on RCP/LCP polarizations of the transmitted light was observed, specifically in the Fano line shape. The origin of such differential response is solely geometric in nature and depends on the orientation angle of the axis of the diattenuating linear retarder (the plasmonic grating) with the direction of incidence of light giving rise to a geometric phase between the plasmonic grating and the waveguide mode [1]. The Fano line shapes were fitted using eq.2, and the asymmetry parameter of the resonance was extracted for both polarization projections shown in **Figure 2b**. A clear difference in the asymmetry of the resonances in both polarization was observed for orientations other than $0°$ and $90°$. In order to quantify such polarization projection dependent asymmetry of the resonance, we defined $\Delta q_{geo.}$ given as

$$\Delta q_{geo.} = abs\left(\frac{q_1 - q_2}{q_1 + q_2}\right)$$

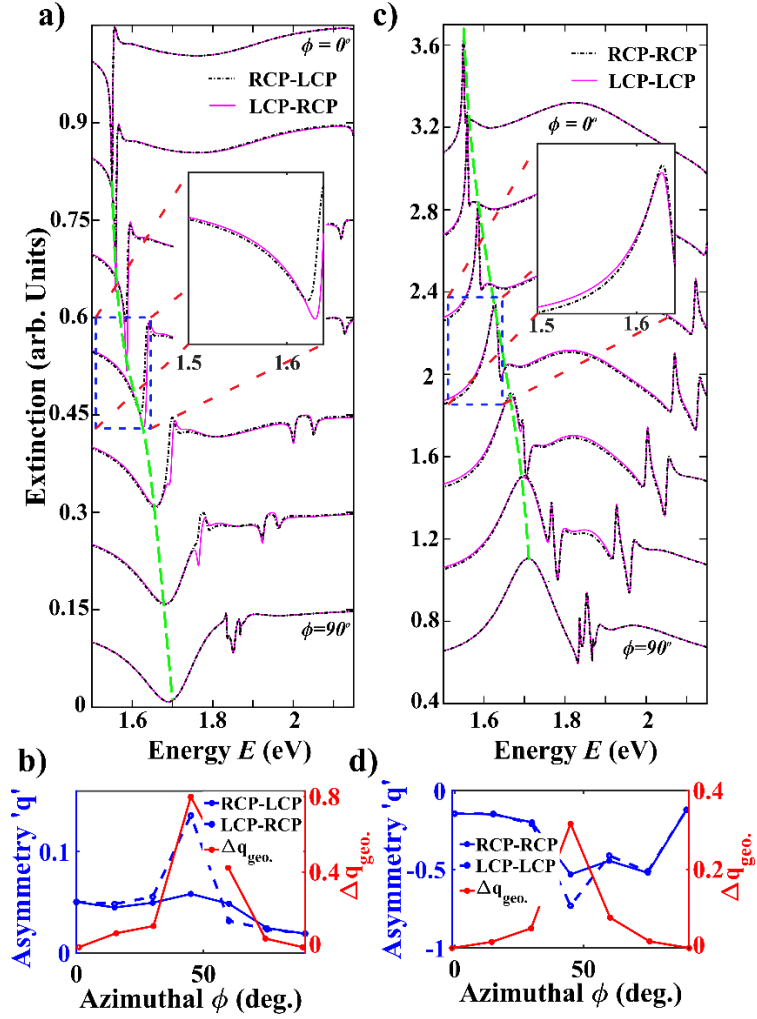

**Figure 2: The dependence of extinction spectra on the azimuthal angle of incidence of a left/right circular polarized light on the grating.** a) Numerically calculated extinction spectra of the grating at inclined angle of incidence with varying azimuthal angle of incidence ($\phi$) from 0° to 90° in step of $15^o$ for (a) cross-polarized projections (RCP projected on LCP and LCP projected on RCP) (c) co-polarized projections (RCP projected on RCP and LCP projected on LCP) are shown. The Fano line shape was fitted using eq.2 and the asymmetry parameter 'q' (left axis) obtained with varying azimuthal angle is shown for both (b) cross-polarized projections and (d) co-polarized projections, a dependence of $\Delta q_{geo.}$ (right axis) with the orientation is also shown. The findings establishes the role of geometric phase on the asymmetry in the extinction spectra.

Here, $q_1$ and $q_2$ are the $q$ values observed for the two circular polarization projections which are being compared. Note that $\Delta q_{geo.}$ depends on the geometry (geometrical orientation) of the anisotropic scattering system as shown in **Figure 2b**. Such dependence of Fano line shape on the azimuthal angle of incidence is a signature of the effect of the geometric phase of light on the Fano line shape with a maximum difference in the asymmetry parameter at $\phi = 45°$ and no difference for $\phi = 0°$ and 90°. It is well known that right and left circularly polarized light acquires equal and opposite Pancharatnam Berry geometric phases while propagating through an anisotropic medium (phase retarder having linear retardance $\delta$), the magnitude of which depends upon the orientation angle ($\beta$) of the anisotropy axis of the medium [23-27]. It has been argued that the geometric phase in such case should be reflected in the crossed circular polarization component. Such arguments stem from empirically decomposing the Jones matrix of a phase retarder as a coherent sum of the Jones matrices of a half-wave retarder ($\delta = \pi$) and an identity matrix. However, we would like to note that such decomposition is strictly valid for a half-wave retarder only. For any arbitrary linear retarder or diattenuating linear retarder, on the other hand, the dependence of the geometric phase on the orientation angle of the axis of the retarder and the projection polarization state is rather complex, as discussed subsequently. Never-the-less, in such case, the geometric phase depends upon the orientation angle of the anisotropic system and may also get reflected in the co-circularly polarized component of light. In **Figure 2c**, we have shown the extinction cross section from the sample for incident LCP/RCP with the light transmitted from the sample being projected on same polarization, i.e., LCP/RCP. It can be noted from the figure that the system shows a differential response for the two polarization projections similar to the earlier case. These results clearly illustrate that indeed a geometrical phase is acquired even in the co-polarized polarization projections in contrast to the model. In what follows, we experimentally demonstrate that the geometric phase can indeed be used for tuning the asymmetry of Fano line shape of spatially tailored waveguided plasmonic crystals having a spatially varying axis of the gratings.

In **Figure 3a** we have shown a schematic of our experimental setup. A dark-field microscope integrated with a polarization state generator (PSG) unit and a polarization state analyzer (PSA) unit. The PSG unit consists of a polarizer (P2) followed by a quarter waveplate (Q2), and these elements are arranged in reverse order for the PSA unit. The PSG and PSA units can be used to generate and analyze an arbitrary state of polarization respectively. The annular illumination of the dark-field microscope was used to record the scattered field spectral response from the spatially tailored waveguided plasmonic crystals (shown in scanning electron microscope (SEM) image) (grating size $10\ \mu m \times 10\ \mu m$, grating periodicity 550 nm, grating rod width 80 nm, height 22 nm; grating separation $20\ \mu m$ (both in x and y directions), the gratings were deposited on $190\ nm$ thick ITO layer). A Nano translation stage was used to move the sample in the focal plane of the microscope such that only a small portion of the plasmonic crystal get illuminated by the light as shown in the inset of **fig3a**. It can be noted from the inset dark field images that the light can be focused such that we get scattering from grating oriented at particular angle (with a particular orientation angle of the axis of diattenuating retarder). The field scattered from the illuminated portion is collected through spectrometer after passing through the PSA unit. The PSA and PSG unit were used to record the scattering spectra for both circular polarization input as well as output combinations with different orientation angle of the grating as shown in **Figure 3**. It can be noted from the figure that a significant amount of change in the asymmetry of line shape is observed with different orientation angle of grating for both co and cross-polarized polarization projections. The $\Delta q_{geo.}$ parameters show that the geometric phase of light indeed modulates the asymmetry of the line shape of the Fano resonance with $\Delta q_{geo.} \sim 0$ for $\phi = 0°$ and $90°$.

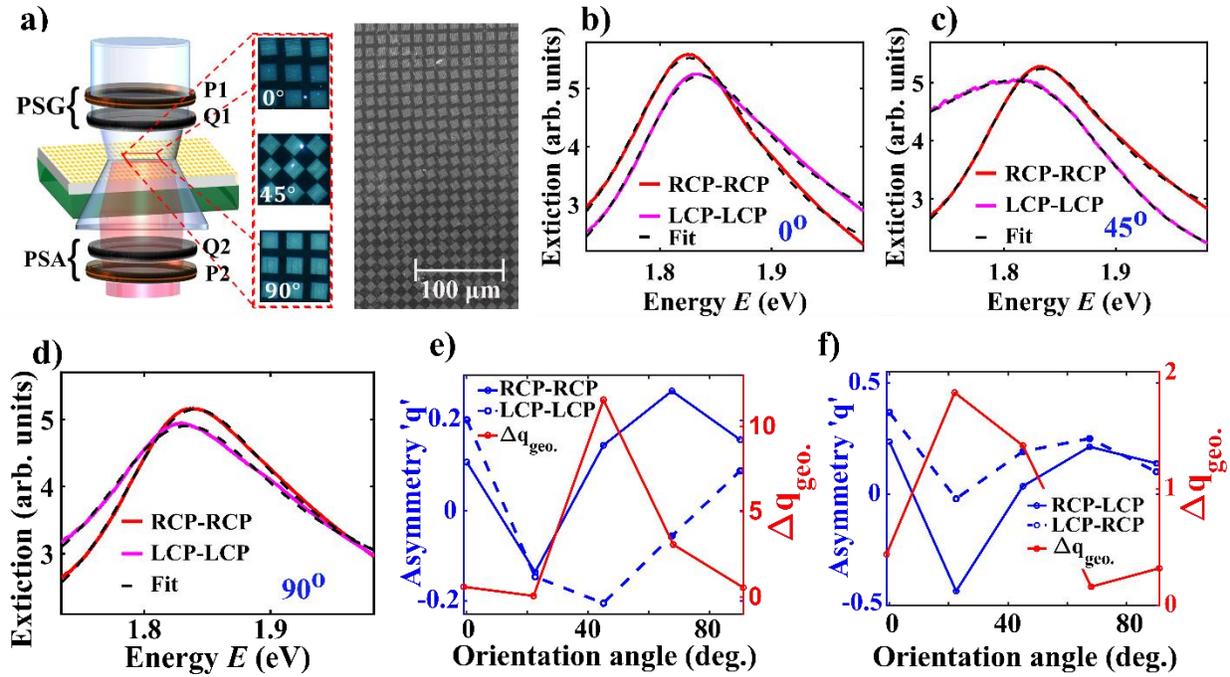

**Figure 3**: **Geometric phase controlled tuning of extinction spectra for spatially tailored waveguided plasmonic crystals**. (a) Schematic of a dark-field microscope integrated with a polarization state generator (PSG) and polarization state analyzer (PSA) unit. The inset shows the dark field images of the plasmonic grating from orientation angles 0°, 45° and 90°, and the SEM image of the spatially tailored waveguided plasmonic crystals. (b-d) The corresponding spectral response of the system for the co-polarized projections is shown respectively. The dependence of the asymmetry parameter 'q' (left axis, blue lines) with the grating orientation angle for (e) the co-polarized projections (LCP projected on LCP and RCP projected on RCP) and (f) the cross-polarized projections (LCP projected on RCP and RCP projected on LCP) are shown, a dependence of $\Delta q_{geo.}$ (right axis) with the orientation is also shown. The dependence of the asymmetry of resonance of the geometrical orientation of the grating clearly shows the role of the geometric phase of light in tuning the Fano phase.

We recorded the full polarization response of the grating for one such axial orientation in the form of the Mueller matrix of the grating sample (**shown in supplementary**). The recorded Mueller matrix was used to extract the linear diattenuation ($d$) and linear retardance ($\delta$) parameters of grating shown in **Figure 4**. It can be noted from the figure that the plasmonic grating system indeed acts as a diattenuating retarder with a small phase retardance value $\delta$ ~0.3 rad and a small linear diattenuation (~0.25) in scattering from the system.

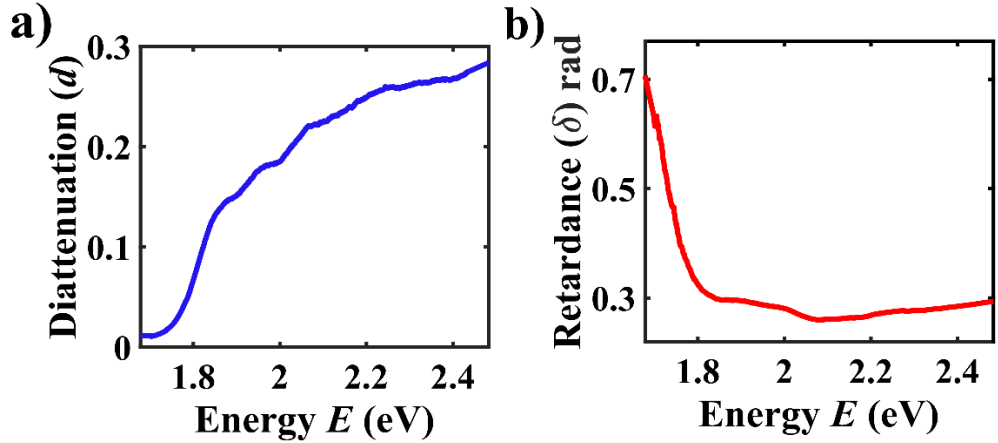

**Figure 4**: **The polarization parameters linear diattenuaton ($d$) and linear retardance ($\delta$) extracted from the Mueller matrix of a waveguided plasmonic grating crystal.** (a) Diattenuation ($d$) parameter showing small amplitude anisotropy ($d \approx 0.25$) and, (b) retardation ($\delta$) parameter showing small phase anisotropy ($\delta \approx 0.3$).

**Discussion**

It is well known that the Panchratnam Berry geometric phase acquired by circularly (or elliptically) polarized light while propagating through an anisotropic medium exhibiting linear retardance effect depends upon the orientation angle of the axis of retarder [23-27]. In the simplest case of a pure half-wave retarder ($\delta = \pi$), the acquired geometric phase ($\phi_g$) has a rather simple dependence on the orientation angle ($\phi_g = 2\beta$) [23-27]. As previously discussed, in such case, empirically it has been shown that the resulting geometrical phase gets accumulated in the cross-polarized component of incident circular polarized light [23-27]. However, for an arbitrary linear retarder ($\delta \neq \pi$), there is no such closed-form analytical expression connecting the geometric phase and the orientation angle. It has been shown that in such case, the polarization evolution may follow some complicated trajectories on the Poincare sphere and the resulting geometric phase has a rather complex dependence on the orientation angle of the axis of the retarder [28-30]. The situation is even more complex for arbitrary diattenuating retarder ($\delta \neq \pi, d \neq 0$), as evident from the results presented in **Fig.2 and 3**, and in such case, the geometric phase information is also encoded in the co-circularly polarized component of the light also. Never-the-less, the geometric phase acquired by left or right circularly polarized light while propagating through arbitrary diattenuating retarder system depends upon the orientation angle of the anisotropy axis. Thus, despite the ensuing complexities on the dependence of the geometric phase on the orientation angle, the results presented in **Fig 2 and 3** clearly demonstrates that the spectral asymmetry of Fano resonance in our diattenuating retarder system of waveguided plasmonic crystal is related with the geometric phase of light that evolves due to the interaction of polarized light with such system. These results validate the physical nature of the phase shift of Fano interference and its connection with the spectral asymmetry of Fano resonance. It also demonstrates that the polarization-tuned geometrical phase of light can be used to modulate the spectral line shape of Fano resonance in an appropriately designed anisotropic Fano resonant systems.

**Conclusion**

In conclusion, we have demonstrated a fundamentally interesting concept associated with Fano resonance by establishing and demonstrating a relationship between the Fano spectral asymmetry parameter *q* and a physical phase shift of Fano interference between a continuum and a narrow resonance mode. Using geometric phase metasurface of waveguided plasmonic crystal, we have experimentally demonstrated control over this phase shift and the resulting asymmetry of Fano spectral line shape through polarization-tuned geometrical phase of light. The experimental results are supported further by the results of FEM based COMSOL simulations. In the process, the results of these studies also uncover new insights and open up discussions on the dependence of geometric phase of light on the orientation of anisotropy axis of arbitrary diattenuating retarder, and on the projection polarization state of light. The results reveal that the Fano *q*-parameter can be controlled by a physically accessible phase of interference, the findings illustrate that the geometrical phase and polarization state of light can be used as an additional degree of freedom to control and manipulate the Fano resonance.

## **References**

1. U Fano. Effects of configuration interaction on intensities and phase shifts. *Physical Review*, **124**(6), 1866, 1961.
2. AC Johnson, Charles M Marcus, MP Hanson, and AC Gossard. Coulomb modified Fano resonance in a one-lead quantum dot. *Physical review letters*, **93**(10), 106803, 2004.
3. I Mazumdar, ARP Rau, and VS Bhasin. Efimov states and their Fano resonances in a neutron-rich nucleus. *Physical review letters*, **97**(6), 062503, 2006.
4. AR Schmidt, MH Hamidian, P Wahl, F Meier, AV Balatsky, JD Garrett, TJ Williams, GM Luke, and JC Davis. Imaging the Fano lattice to'hidden order'transition in uru (2) si (2). *Nature*, **465**(7298), 570, 2010.
5. P Fan, Z Yu, S Fan, and ML Brongersma. Optical Fano resonance of an individual semiconductor nanostructure. *Nature materials*, **13**(5), 471, 2014.
6. B Luk'yanchuk, NI Zheludev, SA Maier, NJ Halas, P Nordlander, H Giessen, and CT Chong. The Fano resonance in plasmonic nanostructures and metamaterials. *Nature materials*, **9**(9), 707, 2010.
7. C Ott, A Kaldun, P Raith, K Meyer, M Laux, J Evers, CH Keitel, CH Greene, and T Pfeifer. Lorentz meets Fano in spectral line shapes: a universal phase and its laser control. *Science*, **340**(6133), 716, 2013.
8. Y Sonnefraud, N Verellen, H Sobhani, GAE Vandenbosch, VV Moshchalkov, PV Dorpe, P Nordlander, and SA Maier. Experimental realization of subradiant, superradiant, and Fano resonances in ring/disk plasmonic nanocavities. *ACS nano*, **4**(3), 1664, 2010.

9. A Christ, SG Tikhodeev, NA Gippius, J Kuhl, and H Giessen. Waveguide plasmon polaritons: strong coupling of photonic and electronic resonances in a metallic photonic crystal slab. *Physical Review Letters*, **91**(18), 183901, 2003.
10. JN Anker, WP Hall, O Lyandres, NC Shah, J Zhao, and RPV Duyne. Biosensing with plasmonic nanosensors. *Nature materials*, **7**(6), 442, 2008.
11. W-S Chang, JB Lassiter, P Swanglap, H Sobhani, S Khatua, P Nordlander, NJ Halas, and S Link. A plasmonic Fano switch. *Nano letters*, **12**(9), 4977, 2012.
12. A Bärnthaler, S Rotter, F Libisch, J Burgdörfer, S Gehler, U Kuhl, and H-J Stöckmann. Probing decoherence through Fano resonances. *Physical review letters*, **105**(5), 056801, 2010.
13. C Wu, AB Khanikaev, R Adato, N Arju, AA Yanik, H Altug, and G Shvets. Fano-resonant asymmetric metamaterials for ultrasensitive spectroscopy and identification of molecular monolayers. *Nature materials*, **11**(1), 69, 2012.
14. K Nozaki, A Shinya, S Matsuo, T Sato, E Kuramochi, and M Notomi. Ultralow-energy and high-contrast all-optical switch involving Fano resonance based on coupled photonic crystal nanocavities. *Optics express*, **21**(10), 11877, 2013.
15. A Kaldun, C Ott, A Blättermann, M Laux, K Meyer, T Ding, A Fischer, and T Pfeifer. Extracting phase and amplitude modifications of laser-coupled Fano resonances. *Physical review letters*, **112**(10), 103001, 2014.
16. MR Shcherbakov, PP Vabishchevich, VV Komarova, TV Dolgova, VI Panov, VV Moshchalkov, and AA Fedyanin. Ultrafast polarization shaping with Fano plasmonic crystals. *Physical review letters*, **108**(25), 253903, 2012.
17. C Wu, AB Khanikaev, and G Shvets. Broadband slow light metamaterial based on a double-continuum Fano resonance. *Physical review letters*, **106**(10), 107403, 2011.
18. B Zhang. Electrodynamics of transformation-based invisibility cloaking. *Light: Science & Applications*, **1**(10), e32, 2012.
19. NI Zheludev, SL Prosvirnin, N Papasimakis, and VA Fedotov. Lasing spaser. *Nature Photonics*, **2**(6), 351, 2008.
20. Y Zhu, X Hu, Y Huang, H Yang, and Q Gong. Fast and low-power all-optical tunable Fano resonance in plasmonic microstructures. *Advanced Optical Materials*, **1**(1), 61, 2013
21. SK Ray, S Chandel, AK Singh, A Kumar, A Mandal, S Misra, P Mitra, and N Ghosh. Polarization tailored Fano interference in plasmonic crystals: A Mueller matrix model of anisotropic Fano resonance. *ACS nano*, **11**(2), 1641, 2017
22. S Pancharatnam. "Generalized theory of interference and its applications." *Proceedings of the Indian Academy of Sciences-Section A*. **44**(6) 1956.
23. S.D. Gupta, N. Ghosh, A. Banerjee *Wave Optics: Basic Concepts and Contemporary Trends*, (CRC Press, 2015)
24. R Bhandari. "Polarization of light and topological phases." *Physics Reports* **281**(1), 1, 1997.

25. ZE Bomzon, G Biener, V Kleiner, and E Hasman. Space-variant Pancharatnam–Berry phase optical elements with computer-generated subwavelength gratings. *Optics letters*, **27**(13), 1141, 2002.
26. L Marrucci, C. Manzo, and D Paparo. Optical spin-to-orbital angular momentum conversion in inhomogeneous anisotropic media. *Physical review letters*, **96**(16), 163905, 2006.
27. X Ling, X Zhou, K Huang, Y Liu, CW Qiu, H Luo and S Wen, Recent advances in the spin Hall effect of light. *Reports on Progress in Physics*, **80**(6), 066401, 2017.
28. D Lopez-Mago, A Canales-Benavides, RI Hernandez-Aranda, and JC Gutiérrez-Vega. Geometric phase morphology of Jones matrices. *Optics letters*, **42**(14), 2667, 2017.
29. JC Gutiérrez-Vega. Pancharatnam–Berry phase of optical systems. *Optics letters*, **36**(7), 1143, 2011.
30. T Van Dijk, HF Schouten, and TD Visser. Geometric interpretation of the Pancharatnam connection and non-cyclic polarization changes. *JOSA A*, **27**(9), 1972, 2010.
31. J Lages, R Giust, and JM Vigoureux, Geometric phase and Pancharatnam phase induced by light wave polarization. *Physica E: Low-dimensional Systems and Nanostructures*, **59**, 6, 2014.
32. J. Soni, H. Purwar, H. Lakhotia, S. Chandel, C. Banerjee, U. Kumar, and N. Ghosh, "Quantitative fluorescence and elastic scattering tissue polarimetry using an Eigenvalue calibrated spectroscopic Mueller matrix system", Optics Express, **21**, 15475, 2013.
33. B. Laude-Boulesteix, A. De Martino, B. Drévillon and L. Schwartz, "Mueller polarimetric imaging system with liquid crystals," Appl. Optics, **43**, 2824, 2004

## Supplementary Information

**Mueller matrix measurement strategy**

A $4 \times 4$ Mueller matrix was constructed using a polarization state generator (PSG) and a polarization state analyzer (PSA) units [32]. The PSG unit consisting of a fixed polarizer (P1) followed by an achromatic quarter-wave plate (QWP1) with the orientation axis kept at $\theta_g$ w.r.t the polarizer (P1). In the same way, a PSA unit was made with an achromatic quarter waveplate (QWP2) with the orientation axis kept at $\theta_a$ followed by a polarizer (P2) kept orthogonal to the polarizer P1. The angle of the orientation angle of quarter wave plates axis were subsequently changed to a set of four optimized angles ($\theta = 35°, 70°, 105°$ and $140°$), such that a total of 16 spectral intensity measurements were done sequentially to record the complete polarization response of the sample integrated with the polarization response of the complete system. The 16 measurements were arranged in particular order to make $4 \times 4$ spectral matrix representing Mueller of the complete system ($M_i$) [32, 33].

$$M_i(\lambda) = A(\lambda) M_s(\lambda) W(\lambda)$$

Here, $A(\lambda)$ and $W(\lambda)$ were made from the four Stokes vectors analyzed and generated from the PSA and PSG units respectively. The Mueller matrix of the sample can be evaluate as

$$M_s(\lambda) = A^{-1}(\lambda) M_i(\lambda) W^{-1}(\lambda)$$

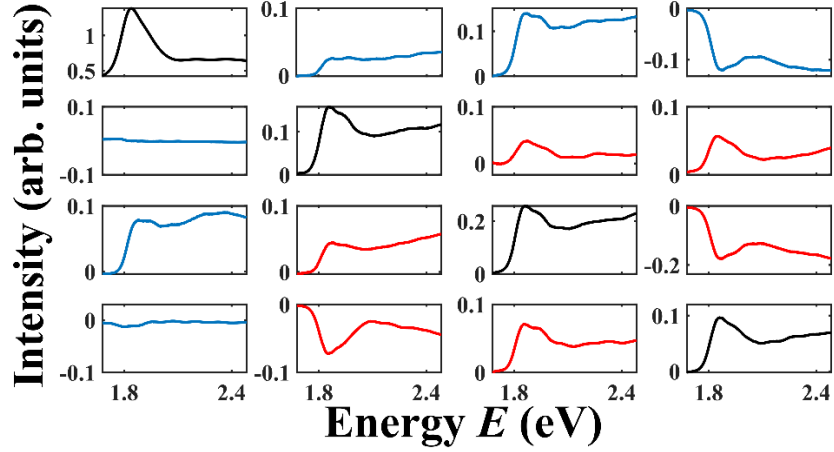

**Figure S1: The $4 \times 4$ Mueller matrix of the plasmonic grating sample encoding the full polarization response of the sample.** The $M_s(1,1)$ element represent the unpolarized intensity of light scattered from the system. The elements marked in blue and red color can be linked with the amplitude and phase anisotropy of the plasmonic grating.

A robust eigenvalue calibration technique was used to extract the wavelength-dependent polarization response of the sample independent of the spectral response of the experimental system response [33]. This strategy was used to record the polarization response from the spatially tailored waveguided plasmonic crystals, the Mueller matrix for one such orientation is shown in figure S1. The matrix shows that the system possesses small amplitude anisotropy associated with the blue lines as well as a small phase anisotropy associated with the elements marked in red. The Mueller matrix of the plasmonic grating and the polarization response extracted from the system establishes that the system is indeed a diattenuating retarder.